\documentclass{article}

\usepackage{amsmath}
\usepackage{graphicx}
\usepackage{comment}
\usepackage{amsfonts}
\usepackage[utf8]{inputenc}
\usepackage{hyperref}
\usepackage{authblk}

\usepackage[numbers]{natbib}                                                     % Ambiente de referencias utilizado.

\newcommand{\R}{\mathbb{R}}

\newcommand{\Z}{\mathbb{Z}}
\newcommand{\N}{\mathbb{N}}

\newcommand{\cur}[1]{\mathcal{#1}}

\newcommand{\minus}{\setminus}

\hypersetup{
  colorlinks   = true, %Colours links instead of ugly boxes
  urlcolor     = blue, %Colour for external hyperlinks
  linkcolor    = blue, %Colour of internal links
  citecolor   = blue %Colour of citations
}

\title{A persistent-homology-based turbulence index \& some applications of TDA on financial markets}

\author[1]{Miguel A. Ruiz-Ortiz} 
\affil[1]{Universidad de Guanajuato, Guanajuato 36240, Gto., M\'EXICO; miguel.ruiz@cimat.mx}

\author[2]{Jesús Rodríguez-Viorato} 
\affil[2]{CIMAT, Jalisco S/N, Col. Valenciana, CP 36023 Guanajuato, Gto, México; jesusr@cimat.mx}

\author[3]{José Carlos Gómez-Larrañaga} 
\affil[3]{CIMAT-Mérida, Carretera Sierra Papacal Chuburna Puerto Km 5, 97302 Sierra Papacal, Yuc., México;  jcarlos@cimat.mx}

\begin{document}

\maketitle

\begin{abstract}
Topological Data Analysis (TDA) is a modern approach to Data Analysis focusing on the topological features of data; it has been widely studied in recent years and used extensively in Biology, Physics, and many other areas. However, financial markets have been studied slightly through TDA. Here we present a quick review of some recent applications of TDA on financial markets, including applications in the early detection of turbulence periods in financial markets and how TDA can help to get new insights while investing. Also, we propose a new turbulence index based on persistent homology --- the fundamental tool for TDA --- that seems to capture critical transitions in financial data; we tested our index with different financial time series (S\&P500, Russel 2000, S\&P/BMV IPC and Nikkei 225) and crash events (Black Monday crash, dot-com crash, 2007-08 crash and COVID-19 crash).  Furthermore, we include an introduction to persistent homology so the reader can understand this paper without knowing TDA. 
\end{abstract}

\begin{center}
\end{center}

\section{Introduction}

The main objective of Topological Data Analysis (\textbf {TDA}) is to find the \textit{shape of data}, checking if there are connected components, holes, or gaps in a representation of data in $\R^d$. This has been achieved mainly through \textbf {persistent homology} \cite{CarlssonTDA, Edels}, and it deals with the detection of $n$-dimensional holes, $ n = 0,1, 2, \ldots $, that are formed by connecting nearby points.

There are at least two good reasons for applying persistent homology to financial data. First, persistent homology gives us shape features of financial data, although it is a high-dimensional space. Second, persistent homology is robust to perturbations of input data, and this robustness is ideal for reliable prediction of regime shifts in markets \cite{otter2017roadmap, PHTI}.

In this paper, we review what kind of efforts have been made to understand financial markets via TDA and how TDA could help us control risk while investing in financial markets. Furthermore, we propose a new persistent-homology-based turbulence index employing previous ideas \cite{GideaCryptos, PHTI} and show experiments results studying this index on different financial datasets during different previous financial crashes (Black Monday, dot-com, 2007-08 and COVID crashes). 

We found that applications in financial markets are in their early stages of development at the time this research was done (see \cite {GideaNetworks, GideaLandscapes, GideaCryptos, Truong, PHTI}). A common way of studying financial markets is through the time series given by the prices of financial assets. As can be seen in \cite {TDATSeries2}, the application of TDA to time series is relatively new and deals with areas such as dynamical systems and signal processing (see \cite {PereaSignalP, Regimes}).

In section \ref{persistencehomology}, we find an introduction to persistent homology ---the most popular tool of TDA --- together with some theoretical concepts to fully understand the applications to financial markets explained here. In general, the main application of TDA in financial markets has been the study of \textbf {critical transitions}, which are the applications on which we focus the most here. We mean by a critical transition an abrupt change in the behavior of a complex system. A critical transition in financial markets is known as a market \textbf {crash}.

In section \ref{previouswork}, we detail three TDA applications to detect critical transitions in financial markets \cite{GideaNetworks, GideaLandscapes, GideaCryptos}. Getting early signs of critical transitions in financial markets would help gain better control of risk in portfolio construction (see section \ref{financial_section} for more details). This, however, is a challenging task, as \cite {GideaNetworks} mentions.

%FIXME: Mofificar la descripción de la última sección.
In section \ref{phtisection}, the development of investing strategies by \cite{PHTI} derived from a persistent-homology-based turbulence index (PHTI) is examined. Finally, in section \ref{section_index}, we propose a modified variation of the PHTI that depends on five parameters. We selected the parameters that best capture the abrupt change in S\&P500 prices due to the 2020 stock market crash between February 20th and April 7th, 2020, induced by the COVID-19 pandemic. We finally tested the selected parameters on different datasets and financial crashes and analyzed the results.

\subsection{Financial markets concepts} \label{financial_section}

A \textbf{financial asset} is an investing vehicle. Some examples of financial assets are shares of stocks, bonds, foreign currencies, and real estate. The three main characteristics of financial assets are liquidity (the capacity to convert the asset into money), profit, and risk. These assets' prices follow the laws of supply and demand, so prices fluctuate over time. A \textbf {portfolio} is simply a collection of financial assets. These assets are acquired in the \textbf{financial markets} (stock market, foreign exchange market, or cryptocurrency market).

We may buy a financial asset, for example, shares of a company, with the goal in mind of obtaining a profit. This can be achieved if the price of this financial asset increases after we have bought it and then we sell it. Transactions of financial assets have the risk of having losses instead of positive returns (a return is a change in the price of an asset over time) because the price may decrease after we have bought it.

Due to the volatility and uncertainty that often exist in financial markets, it is important to have risk measures to avoid large losses caused by extreme market movements. This topic has been widely studied with statistics and artificial intelligence tools (see \cite {mcneil2015, trippi1995}).

\section{Persistent Homology}\label{persistencehomology}

This section introduces the TDA concepts necessary to understand this paper. To go deeper, see \cite {CarlssonTDA, Edels, CompuTopo}.

\subsection {Simplicial Homology}

\begin {figure} [h]
  \centering
    \includegraphics [scale = 0.23] {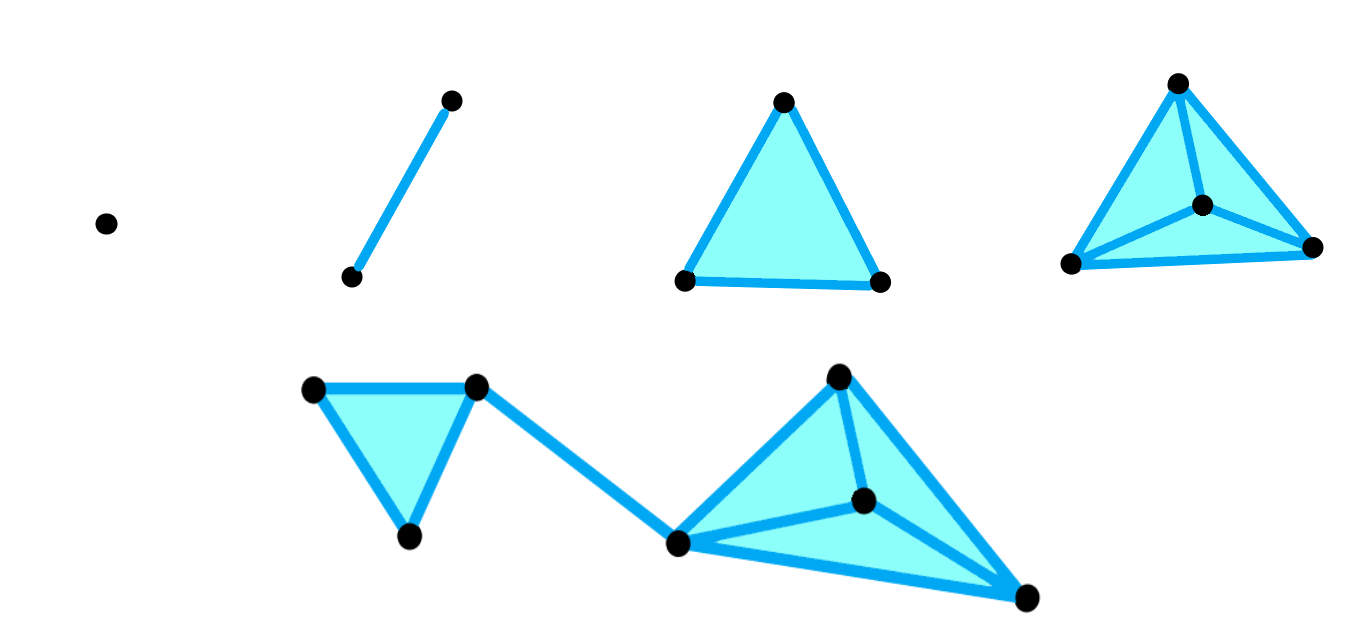}
  \caption {$ k $ -simplices for $ k = 0, 1, 2, 3 $ (top) and a simplicial complex (bottom).}
  \label {img1}
\end {figure}

\textbf {Definition:} A \textbf {simplicial complex} $ \Sigma $ is a collection of non-empty subsets of a finite set of vertices $ \Sigma_0 $ that satisfy
\[
\alpha \in \Sigma, \beta \subset \alpha \quad \Rightarrow \quad \beta \in \Sigma.
\]
If $ \alpha \in \Sigma $ and $ k = |\alpha| -1 $, $ \alpha $ is called a $ k $ - \textbf {simplex}, and $ \beta \subset \alpha $ is known as a \textbf {face} of $ \alpha $. We will denote the set of all $ k $ -simplices as $\Sigma_k $. The \textbf {dimension of} $ \Sigma $ is defined as
\[
\dim (\Sigma): = \max_{\alpha \in \Sigma} \{|\alpha| -1 \}.
\]
These concepts can be visualized in figure \ref {img1}. In TDA, simplices are used to study data.

Let us consider $ C_k $ the free abelian group with coefficients in $\Z$ generated by the $ k $-simplices, so each element $ c \in C_k $ can be expressed as
\[
c = \sum_{i=1}^n c_i \alpha_i,
\]
where the $ \alpha_i $ are $ k $-simplices and $ c_i \in \Z $, for some $n\in \N$.

Suppose that a total order is imposed on the set of vertices $ \Sigma_0 $. This allows defining a set of homomorphisms $ d_i: C_k \to C_{k-1} $, $ 0 \leq i \leq k $, given by their action on the basis $\Sigma_k$ as $ d_i (\sigma) = \sigma \minus \{s_i \} $ for each $\sigma\in \Sigma_k$ (i.e., we first define it on the basis $\Sigma_k$ and then extend it linearly to $C_k$), where $ s_i $ is the $ i $ -th element of $ \sigma$. Let us define the \textit {boundary homomorphism} $ \delta_k: C_k \to C_ {k-1} $ given by
\[
\delta_k = \sum_ {i = 0} ^ k (-1) ^ i d_i.
\]

Notice that $ \delta_k \circ \delta_ {k + 1} = 0 $, so $ Im (\delta_ {k + 1}) \subseteq Ker (\delta_k) $. This leads to the following definition.

\textbf {Definition:} The $k$-\textit {th homology group of} $ \Sigma $ is defined as
\[
H_k (\Sigma) = Ker (\delta_k) / Im (\delta_ {k + 1}).
\]
The $k$-\textit {th Betti number} of $ \Sigma $ is defined as
\[
\beta_k = rank (H_k (\Sigma)),
\]
and it is the number of $k$-dimensional holes on $\Sigma$, e.g., $ \beta_0 $ is the number of independent connected components, $ \beta_1 $ is the number of loops, $ \beta_2$ is the number of cavities, and so on.

\subsection{Construction of simplicial complexes}
Suppose we have a point cloud $ X = \{x_1, x_2, \ldots, x_N \} \subset \R ^ d $. To study its shape, we would like to associate $X$ with a simplicial complex. The following are some different ways to do it.

\subsubsection {Ĉech simplicial complex}
\textbf {Definition:} The \textit {Ĉech complex} $ C (X, \epsilon) $ with parameter $ \epsilon> 0 $ and with vertices in $ X $ is
\[
C (X, \epsilon) = \left \{\alpha \subset X \,: \, \bigcap_ {v \in \alpha} B_ \epsilon (v) \neq \emptyset \right \}
\]
where $ B_ \epsilon (v) $ is the ball of radius $ \epsilon $ centered at $ v $. This simplicial complex is called the \textit {nerve} of $ \{B_ \epsilon (x) \} _ {x \in X} $.

\subsubsection {Vietoris-Rips simplicial complex}

\textbf {Definition:} The \textit {Vietoris-Rips complex} $ VR (X, \epsilon) $ with parameter $ \epsilon> 0 $ and with vertices in $ X $ is
\[
VR (X, \epsilon) = \left \{\alpha \subset X \,: \, || v_i-v_j || \leq \epsilon, \, \forall \, v_i, v_j \in \alpha \right \}.
\]

Notice that if $ \Sigma (\epsilon) $ is the Ĉech complex or the Vietoris-Rips complex associated with $ \epsilon> 0 $ with vertices at $ X $, we get
\[
\Sigma (\epsilon ') \subseteq \Sigma (\epsilon), \quad \forall \; \epsilon '\leq \epsilon.
\]
We call this the \textbf {filtering property}. The inclusion map induces a morphism between the homology groups:
\[
H_k (\Sigma (\epsilon ')) \rightarrow H_k (\Sigma (\epsilon)), \quad \forall \; \epsilon '\leq \epsilon.
\]
These last two results are helpful for the definition of persistent homology, as we will see next.

%\subsubsection{Clique simplicial complex} \label{clique-complex}
%
%Sometimes the data is presented through a graph $ G = (V, E) $, where $ V $ is the set of vertices and $ E $ is the set of edges. An example of this is the correlation networks used in \cite {GideaNetworks} (see section \ref{correlation_graphs}).

%\textbf {Definition:} Given a graph $ G = (V, E) $, a \textit {clique} $ C $ is a set of vertices of $ V $ such that every pair of different vertices in $C$ are connected by an edge in $ E $.

%With the previous definition, it is now possible to define the clique complex (see \cite {CliqueComplex}).

%\textbf {Definition:} Let $ G = (V, E) $ be a graph. The \textit {clique complex} $ X (G) $ of $ G $ is the simplicial complex resulting from considering each clique with $k$ vertices as a $ (k-1) $-simplex.

%Suppose there is a function $ w: E \to [0, \infty) $ that assigns a weight to each edge of $ G $ ---i.e., $ G $ is a \textit {weighted graph}. Let us consider $ \theta_{max} = \max (w) $ and $ \theta \in [0, \theta_ {max}] $. Let $ G (\theta) $ be the subgraph of $ G $ that keeps all edges with weights less than or equal to $ \theta $. Note that if $ 0 \leq \theta '\leq \theta $, then $ G (\theta') \subset G (\theta) $ (the former is a subgraph of the latter). Even more,
%\[
%X (G (\theta ')) \subseteq X (G (\theta)),
%\]
%that is, the filtering property is obtained.

\subsection{Persistent homology}

\subsubsection{Filtration}

\textbf {Definition:} Let $ \Sigma $ be a simplicial complex. A \textit {filtration} of $ \Sigma $ is a mapping $ \epsilon \in (0, \infty) \mapsto \Sigma (\epsilon) $, where $ \Sigma (\epsilon) \subseteq \Sigma $ is a sub-simplicial complex that fulfills the \textit {filtering property}:
\[
\Sigma (\epsilon ') \subseteq \Sigma (\epsilon), \quad \forall \; \epsilon '\leq \epsilon.
\]

Observe that the Vietoris-Rips and Chech complexes are filtrations. We have already commented that by taking the $k$-dimensional Homology of a filtration $ \epsilon \mapsto \Sigma (\epsilon)$ we obtain induced homomorphism:
\[
h_k^{\epsilon ', \epsilon}: H_k (\Sigma (\epsilon ')) \rightarrow H_k (\Sigma (\epsilon)), \quad \forall \; \epsilon' \leq \epsilon.
\]
The inclusion $\Sigma (\epsilon ') \subset \Sigma (\epsilon)$ induces the map $h_k^{\epsilon ', \epsilon}$. 

\textbf {Definition:} The $ k $\textit {-th persistent homology groups} are given by $ H_k ^ {\epsilon ', \epsilon} = Im (h_k ^ {\epsilon', \epsilon}) $, $ \forall \; \epsilon '\leq \epsilon $.

\textbf {Definition:} We say that the homology class $ \alpha \in H_k (\Sigma (b)) $, $ b> 0 $, \textit {is born} at $ b $ if $ \alpha \not \in H_k ^ {b- \delta, b} $ for all $ \delta> 0 $. If $ \alpha $ is born at $ b $, we say that \textit {dies} at $ d \geq b $ if $ h_k ^ {b, \epsilon} (\alpha) \neq 0 $ for $ b <\epsilon < d $, and $ h_k ^ {b, d} (\alpha) = 0 $. If $ \alpha $ is born and never dies in the above sense, we say that \textit {dies} at $ d = \infty $.

We then have a birth value $ b (\alpha) = b $ and a death value $ d (\alpha) = d $ for each $ \alpha$ in the homology groups.

\textbf {Definition:} The \textit {persistence}, or \textit {lifetime}, of the homology class $ \alpha $ is the difference
\[
pers (\alpha) = d (\alpha) -b (\alpha).
\]

The intuition behind these concepts is that through a ``discrete" filtration of simplicial complexes associated with a finite number of parameters $ 0 <\epsilon_1 <\epsilon_2 <\ldots <\epsilon_n $ we can track the topological characteristics (connected components, holes, etc.) that appear throughout the filtration. Those generators that have more persistence are the most significant.

\subsubsection {Persistence Diagrams}
It is possible to find a set of bases, one for each homology group, such that the image of a base element $\sigma_k$ for $H_k(\Sigma(\epsilon))$ under $h_k^{\epsilon, \epsilon'}$ is either a generator of $H_k(\Sigma(\epsilon'))$ or zero.  

\textbf {Definition:} Let us fix a basis for each homology group as explained above. The $ k $ \textit {-th persistence diagram} (or $ k $ -dimensional persistence diagram) of the filtration $ \epsilon \mapsto \Sigma (\epsilon) $ is defined as the multiset $ P_k $ in $ \R ^ 2 $ obtained as follows:
\begin{itemize}
    \item Each generator class $ \alpha_k $ in a $ k $ -th homology group is assigned a point $ (b_k, d_k) \in \R ^ 2 $ with multiplicity $ \mu_k (b_k, d_k) $, where $ b_k = b (\alpha_k) $ and $ d_k = d (\alpha_k) $ (the birth and death parameters), and the multiplicity is the number of generators $ \alpha_k $ that were born at $ b_k $ and died at $ d_k $.
    \item $ P_k $ contains all points $(x,x)\in \R ^ 2 $ with $x\geq 0$ with infinite multiplicity.
\end{itemize}
%TODO: buscar referencia para la independencia de las bases
Notice the need for bases in the definition of the persistent diagrams. Fortunately, the resulting persistence diagram is independent of the selected bases.

\subsubsection{Distance between persistence diagrams}

We can define many distance functions to make the space $ \cur {P} $ of persistence diagrams a metric space. A very common metric is the following.

\textbf {Definition:} The \textit {Wasserstein distance} of degree $p$, $p\in \N$, is defined as
\[
W_p (P_k ^ 1, P_k ^ 2) = \inf _ {\phi} \left [\sum_ {q \in P_k ^ 1} || q- \phi (q) || ^ p_ \infty \right] ^ { 1/p},
\]
where the infimum is taken over all bijections $ \phi: P_k ^ 1 \to P_k ^ 2 $, and $ || \cdot || _ \infty $ denotes the supremum norm at $ \R ^ 2 $. When $ p = \infty $, the Wasserstein distance $ W_ \infty $ is known as the \textit {bottleneck distance}.

An important property that makes persistent homology suitable for analyzing data with noise is its robustness under small disturbances in the data. That is if our point cloud changes `a little', then the Wasserstein distance between the respective persistence diagrams is small (see \cite{EstabilidadDiagPers}).

\subsection {Persistence landscapes}

The metric space $ (\cur {P}, W_p) $ of persistence diagrams is not complete, so it is unsuitable for statistical purposes. This is why \textit {persistence landscapes} are a preferred representation of persistence diagrams, they are elements in the Banach space $ L ^ p (\N \times \R) $ (recall that a complete normed vector space is called a Banach space), in which statistical methods can be used (see \cite {Landscapes}).

Let us consider $ P $ a persistence diagram. For each point $ (b_ \alpha, d_ \alpha) \in P $ associated to the birth and death of some homological class $ \alpha $, define
\[
f _ {(b_ \alpha, d_ \alpha)} (x) = \begin {cases}
x-b_ \alpha, & x \in \left (b_ \alpha, \frac {b_ \alpha + d_ \alpha} {2} \right]; \\
-x + d_ \alpha, & x \in \left (\frac {b_ \alpha + d_ \alpha} {2}, d_ \alpha \right); \\
0, & x \not \in (b_ \alpha, d_ \alpha).
\end {cases}
\]
Let $ \lambda = (\lambda_k) _ {k \in \N} $ be a sequence of functions $ \lambda_k: \R \to [0,\infty) $ given by
\[
\lambda_k (x) = k \mbox {-} \max \{f _ {(b_ \alpha, d_ \alpha)} (x) \,: \, (b_ \alpha, d_ \alpha) \in P \} ,
\]
where $ k $-$ \max $ denotes the $ k $ -th largest value of a set. When the $ k $ -th largest value does not exist, we set $ \lambda_k (x) = 0 $.

\textbf {Definition:} The \textit {persistence landscape} associated with the persistence diagram $ P $ is the sequence of functions $ \lambda = (\lambda_k) _k \in L ^ p (\N \times \R) $ defined above. $ \lambda_k $ is known as the $ k$-th layer of the persistence landscape.

The norm that we consider in the Banach space $ L ^ p (\N \times \R) $ is given by
\[
|| \eta || _p = \left (\sum_ {k = 1} ^ \infty || \eta_k || _p ^ p \right) ^ {1 / p},
\]
where $ \eta = (\eta_k) _k $ and $ || \cdot || _p $ denotes the norm $ L ^ p $, i.e.,
\[
|| f ||_p = \left (\int_ \R | f | ^ p \right) ^ {1 / p}.
\]

\section{Detecting critical transitions}\label{previouswork}

\subsection{Correlation graphs of Dow-Jones index} \label{correlation_graphs}

M. Gidea in \cite{GideaNetworks} studied the daily returns of company stocks in the Dow Jones Industrial Average (DJIA) during the period January 2004 to September 2008 and tracks the change in the topology of the \textbf {correlation graphs}, defined below, as prices approached the 2007-08 financial crisis using persistent homology. Their objective was to find early signs of such a critical transition ---a crash--- in financial markets.

Let us consider $ x_i (t) $ the daily return $ t $ of the $ i $ -th share. The graph $ G = (V, E) $ of correlation at time $ t $ of the returns $ \{x_i (t) \} _ {t, i} $ is a weighted graph where the stocks of the DJIA are represented by the vertices $ V $, and each pair of distinct vertices $ i, j \in V $ are connected by an edge $ e \in E $, which is assigned the weight
\[
d_ {i, j} (t): = \sqrt {2 (1-c_ {i, j} (t))},
\]
where
\[
c_ {i, j} (t) = \frac {\sum _ {\tau = t-T} ^ t (x_i (\tau) - \overline {x} _i) (x_j (\tau) - \overline {x} _j )} {\sqrt {\sum _ {\tau = t-T} ^ t (x_i (\tau) - \overline {x} _i) ^ 2} \sqrt {\sum_ {\tau = t-T} ^ t (x_j (\tau) - \overline {x} _j) ^ 2}},
\]
is the Pearson correlation coefficient between nodes $ i $ and $ j $ at time $ t $ with a time window $ T $, and $ \overline {x} _i, \overline {x} _j $ are the average of the returns of the action $ i $ and $ j $ in that time window. The author used $ T = $ 15.

%According to the section \ref {clique-complex}
For each $ t $ we can compute the persistent homology of the correlation graph at time $ t $ and obtain information about its topology. The author computed the persistent homology in dimensions 0 and 1. His results showed how there was less correlation between stocks before the 2008 financial crash, and the correlation between stocks increased days before the crisis. This was quantified with the Wasserstein distances between the 0-dimensional persistence diagrams at time $ t $ and at an initial time $ t_0 $. In this time series of distances, the author found greater oscillations days before the crash, thus having signs of the critical transition.

\subsection{Persistence landscapes of financial crashes}

M. Gidea and Y. Katz in \cite {GideaLandscapes} studied the daily returns of the S\&P500, the DJIA, NASDAQ, and the Russel 2000 ---the 4 major financial indices in the United States--- during the technology crash of 2000 and the financial crisis of 2008. Each 1-dimensional time series together form a 4-dimensional time series. The authors used persistent homology to detect topological patterns in this multidimensional time series. Persistence landscapes and their norms $ L ^ p $, $ p = 1, 2 $, were used to measure changes in 1-dimensional persistence diagrams. In their results, they found that the $ L ^ p $ norms have grown considerably days before the two crashes, while the norms behave calmly when the market is stable. The methodology they used is described below.

Let $ \{x_n ^ k \} _ n $, $ k = 1, \ldots, d $, be $ d $ time series and let us fix a window size $ w $. For each time $ t_n $, we have a point $ x (t_n) = (x_n ^ 1, \ldots, x_n ^ d) \in \R ^ d $. Let us consider the matrix
\[
X_n: = [x (t_n) ^ T, x (t_ {n + 1}) ^ T, \ldots, x (t_ {n + w-1}) ^ T]
\]
of size $ d \times w $, which represents a cloud of $ w $ points in $ \R ^ d $. Then, we obtain a filtering of Rips simplicial complexes to obtain the persistence diagram of dimension 1 and then its associated persistence landscape together with its norm $ L ^ 1 $ and $ L ^ 2 $. This is done for each time $ t_n $, obtaining a time series of the norms of persistence landscapes that tracks changes in the topological properties of each point cloud $X_n$ over time.

The authors applied the methodology described before to the daily log-returns of the S\&P500, the DJIA, the NASDAQ, and the Russel 2000 from December 23, 1987, to December 8, 2016 (7301 daily data). Their results showed more persistent 1-cycles emerging in the point clouds when there is more volatility in the market. To quantify this behavior, the $ L^1 $ and $ L^2 $ norms of the persistence landscapes were computed for each point cloud. The set of values obtained induces a daily time series with these quantities and found these norms grow when periods of high volatility in the markets approach, such as the crash of 2000 and 2008. To see the variability of both time series, the author proceeds to calculate the 500-day moving variance. As expected, the time series of the moving variance also grows closer to both financial crashes.

\subsection{Crashes in cryptocurrency market}

In \cite {GideaCryptos}, M. Gidea et al. analyzed the time series of four of the most important cryptocurrencies: Bitcoin, Ethereum, Litecoin, and Ripple ---which made up 66 \% of the total cryptocurrency market capitalization--- using a similar approach to \cite {GideaLandscapes}. The difference between both approaches is that the methodology of \cite {GideaCryptos} was applied to a specific time series, while in \cite {GideaLandscapes} the methodology was applied to $d$ time series. Furthermore, \cite{GideaCryptos} took into account what they define as the $ C^1 $ norm of persistence landscapes and used the $k$ -means algorithm of machine learning, which is a clustering algorithm, to identify patterns in prices and the $C^1$ norms indicating critical transitions before the \textit {cryptocurrency crash} in January 2008, which started with a Bitcoin price drop of 65\%.

The methodology of \cite{GideaCryptos} was the following. Let us consider a time series $ X = \{x_0, \ldots, x_ {N-1} \} $ and set $ d \in \N $. Let
\begin {align*}
    z_t: = (x_t, x_ {t + 1}, \ldots, x_ {t + d-1}) \in \R ^ d, 
\end {align*}
for $t \in \{0, \ldots, N-d \}$. We now have a sequence of point clouds in $ \R ^ d $
\[
Z ^ t: = \{z_t, z_ {t + 1}, \ldots, z_ {t + (w-1)} \}, \quad t \in \{0, \ldots, N-d-w + 1 \},
\]
where $ w $ is the size of a rolling window (the number of points in each cloud). This association between the time series and the sequence of point clouds is known as the \textbf {time-delay coordinate embedding} or \textbf{Takens' embedding}, which is inspired by Takens' theorem (see \cite {Takens}).

Let us fix $ t \in \{0, \ldots, N-d-w + 1 \} $ and consider the point cloud $ Z^t $. We obtain a filtration of Vietoris-Rips simplicial complexes to obtain the 1-dimensional persistence diagram, its persistence landscape $ \lambda ^ t $, and the norm $ L ^ 1 $ of this latter. In this way, we obtain a time series $ \{|| \lambda ^ t || _1 \} _ {t}$.

To detect the cases when $ || \lambda ^ t || _1 $ is large or when there are large jumps between $ || \lambda ^ t || _1 $ and $ || \lambda ^ {t-1} || _1 $, the authors define the \textit{norm} $ C ^ 1 $ of the persistence landscape $ \lambda ^ t $ as
\[
|| \lambda ^ t || _ {C ^ 1} = || \lambda ^ t || _1 + ||| \lambda ^ t || _1- || \lambda ^ {t-1} || _1 |,
\]
which is a definition motivated by the $ C ^ 1 $ norm of differentiable functions. Notice that $||\cdot ||_{C^1}$ is not a norm but just a quantity to identify the patterns of $||\lambda^t||_1$ of interest.

This series of steps were applied to the daily log-returns of Bitcoin, Ethereum, Litecoin, and Ripple from September 14th, 2017, until the most significant previous peak to the critical transition in each cryptocurrency. They took $ d = $ 4 and $ w = $ 50 in their experiments.

The $k$-means algorithm was applied to the points $ \{(x_t, y_t) \} _ t $, where $ x_t $ is the logarithm of the price at time $ t $ of a cryptocurrency and $ y_t = || \lambda ^ t || _ {C ^ 1} $ is the norm $ C ^ 1 $ of the persistence landscape $ \lambda ^ t $. Both time series $ \{x_t \} _ t $ and $ \{y_t \} _ t $ were scaled to the interval $ [0,1] $ prior to this algorithm. $ k $ was chosen using the \textit {elbow method} q

A cluster given by the $k$ -means algorithm was defined as a \textbf {early warning signal} if the cluster contained points $ (x_t, y_t) $ with $ y_t> 0.5 $ and the times $ t $ formed an almost contiguous time interval. If most of the points have $ y_t> 0.5 $, it was a \textbf {strong early warning signal}; and if only a few points had $ y_t> 0.5 $, it was a \textbf {weak early warning signal}.

For the cryptocurrencies Bitcoin, Ethereum and Ripple, early warning signals were obtained prior to the crashes of each cryptocurrency, while for Litecoin the methodology did not work and obtained a false positive. The authors tested their methodology with 8 other different time series, and only 2 had false positives. This indicates to us that it would be worthwhile to research further into the reliability of this methodology to detect early signs of critical transitions in financial markets.

This article was published in May 2019 by Elsevier, and the author mentioned that \cite {GideaNetworks}, \cite {GideaLandscapes}, and \cite {Truong} were the few research papers about TDA for financial data analysis. The major advantage of studying critical transitions in financial markets with TDA is that it is free of statistical assumptions in the time series.

\section{Investing strategies based on a TDA indicator}\label{phtisection}

In the financial asset management industry, it is known that turbulent periods in financial markets are characterized by low returns, high volatility, and the correlation between assets increases. This makes diversification ---the golden rule in investing --- not working as it should. Recall that diversification is an investment strategy made up of a portfolio with a wide variety of distinct investment vehicles (not putting all the eggs in one basket).

One way to deal with turbulent periods while investing is the estimation of the dynamic regime of financial markets ---the behavior that the market has over time--- or the creation of risk indicators that predict changes in the market regime. Both approaches allow us to create investment strategies with low risk and a higher return compared to the \textit {buy-and-hold strategy} \cite {PHTI}.

\cite {PHTI} designed three investment strategies via a persistent-homology-based turbulence index (PHTI) they created. In their research, they simulated these strategies using the PHTI index, the VIX index, and the \cite {Chow} index based on the \textit{Mahalanobis} distance ---the two latter being known turbulence indices. The authors found that the PHTI-based strategies were better according to some performance measures described later.

The PHTI is constructed as follows. Let us fix $N$ industrial portfolios and consider a rolling window of 60 trading days. This leads to a $N$-dimensional point cloud of size 60 where each point represents the daily returns of the $ N $ industrial portfolios in some day. Then, we obtain the 1-dimensional (or 0-dimensional) persistence diagram via a Vietoris-Rips simplicial complex filtration.

Moving the rolling window, we obtain a set of persistence diagrams $ \{P_t \}_t $, and then construct a sequence $ \{W_t \}_t $, where $ W_t $ is the Wasserstein distance between $ P_t $ and $ P_{t-1} $. $ \{W_t \}_t $ is called the PHTI \textit {raw} index. The sequence $ \{W_t \}_t $ is smoothened with a moving average of 60 days and gives rise to the \textbf{PHTI} index $ \{w_t \}_t $:
\[
w_t = \frac{1}{60} \sum_{j = 0}^{59} W_{t-j}.
\]

\cite{PHTI} took into account in the investment strategies the last 60 monthly values of the PHTI. Depending on which quintile the last value falls into, it was the total value of equity in the S\&P500. The following were the three investment strategies the authors designed:
\begin{itemize}
    \item [1)] \textbf {Protection against extreme values of the index.} Investing of the 100 \% during the following month unless the last value falls into the last quintile. In this last case, do not invest anything in that month.
    
    \item [2)] \textbf {Flexible exposure.} If the last value falls in the $ n $ -th quintile, $ n \in \{1, 2, \ldots, 5 \} $, then invest the $ ( 100-20 (n-1)) $ \% of equity (100 \%, 80 \%, 60 \%, 40 \% and 20 \% for $n=1, 2, 3, 4, 5$, respectively).
    
    \item [3)] \textbf {Exposure with leverage.} Leverage allows us to use debt to increase the amount of money for investment. If the last value falls into the $ n $ -th quintile, $ n \in \{1, 2, \ldots, 5 \} $, then invest the $ (120-5n (n-1)) $ \% of equity (120 \%, 110 \%, 90 \%, 60 \% and 20 \% for $n=1,2,3,4,5$, respectively).
\end{itemize}

\cite{PHTI} called this methodology the 60-60-60 rule because of the rolling window size, the moving average, and the 60 monthly values of $ w_t$ considered for the investment strategies. For the other two turbulence indices, the 60-60-60 rule was adapted. 

The authors in their simulations used $ N = 10 $ and $ N = 30 $ industrial portfolios from different sectors to represent the general market situation. We denote the PHTI that uses the 10 and 30 portfolios as PHTI(10) and PHTI(30), respectively. Analogously with the index of \cite {Chow}, denoted here by Chow(10) and Chow(30). This leads to 5 different turbulence indices: PHTI(10), PHTI(30), Chow(10), Chow(30) and VIX. Each of these indices can be used for the 3 different investment strategies described. These strategies were tested from June 1991 to June 2019, encompassing 337 monthly returns of the S\&P500, and were compared to the buy-and-hold strategy as \textit {benchmark}.

The authors used, as performance measures for any investment strategy, the mean annualized return $ \mu $, the mean annualized standard deviation of returns $ \sigma $, the annualized Sharpe ratio \textit {SR} (\textit {Sharpe ratio}, quotient between the two previous measures), and the maximum \textit {drawdown} \textit {maxDD} (maximum drop experienced in the portfolio since the last maximum until this maximum is exceeded again).

In table \ref{table5}, the 4 performance measures are reported. We can see that the PHTI(10) and the PHTI(30) are the indices that showed better Sharpe Ratios and lower maximum drawdowns in the 3 different strategies. In addition, they also perform better than the benchmark with respect to Sharpe Ratio and maximum drawdown. So the PHTI has the potential to be a good turbulence index.

\begin{table*}
\begin{center}
\begin{tabular}{c|cccccc}
\hline
\multicolumn{7}{c}{\textbf{Protection against extreme index values}} \\
\hline 
& Benchmark & PHTI(10) & PHTI(30) & Chow(10) & Chow(30) & VIX  \\
\hline
$\mu$ & 8.55 & 8.94 & 8.87 & 7.27 & 6.35 & 5.49 \\
$\sigma$ & 14.18 & 10.74 & 10.84 & 11.92 & 11.77 & 10.04 \\
\textit{SR} & 0.60 & 0.83 & 0.82 & 0.61 & 0.54 & 0.55 \\
\textit{maxDD} & 52.56 & 33.42 & 33.42 & 41.09 & 44.43 & 39.47 \\
\hline
\multicolumn{7}{c}{\vspace*{0.25cm}} \\
\hline
\multicolumn{7}{c}{\textbf{Flexible exposure}} \\
\hline 
&  Benchmark & PHTI(10) & PHTI(30) & Chow(10) & Chow(30) & VIX  \\
\hline
$\mu$ & 8.55 & 5.99 & 5.96 & 4.58 & 4.60 & 5.10 \\
$\sigma$ & 14.18 & 8.32 & 8.53 & 9.02 & 8.81 & 7.98 \\
\textit{SR} & 0.60 & 0.72 & 0.70 & 0.51 & 0.52 & 0.64 \\
\textit{maxDD} & 52.56 & 24.75 & 26.43 & 35.98 & 35.97 & 35.00 \\
\hline
\multicolumn{7}{c}{\vspace*{0.25cm}} \\
\hline
\multicolumn{7}{c}{\textbf{Exposure with leverage}} \\
\hline 
&  Benchmark & PHTI(10) & PHTI(30) & Chow(10) & Chow(30) & VIX  \\
\hline
$\mu$ & 8.55 & 8.12 & 8.28 & 6.46 & 6.31 & 6.26 \\
$\sigma$ & 14.18 & 10.82 & 11.26 & 11.82 & 11.68 & 10.18 \\
\textit{SR} & 0.60 & 0.75 & 0.74 & 0.55 & 0.54 & 0.62 \\
\textit{maxDD} & 52.56 & 33.15 & 33.28 & 44.08 & 45.06 & 43.52 \\
\hline
\end{tabular}
\caption{Performance measures of the different investment strategies with the different indices. PHTI(10) and PHTI(30) are the persistent-homology-based indices that showed better Sharpe Ratios and lower maximum drawdowns.}
\label{table5}
\end{center}
\end{table*}

\newpage

\section{A new version of Persistent Homology Based Index} \label{section_index}

%DONE: Introduction and explaining the construction
Combining ideas of \cite{PHTI, GideaCryptos, GideaLandscapes}, we propose another turbulence index based on persistent homology.
The construction is the following. Let $\{x_t\}_t$ be a time series of the log-returns of some financial asset. Through Takens's embedding, we obtain a set of $d$-dimensional points
\[
z_t := (x_t, x_{t+\tau}, \ldots, x_{t+\tau(d-1)})\in \R^d,
\]
where $\tau$ is called the \textit{time delay} of the embedding. This is a deviation from \cite{GideaCryptos}, the authors only considered $\tau=1$ and $d=4$. We decided to try with more values, in this case, we tried with $\tau=1,2, 5$ and $d = 3, 4, 5,10$ (see Table \ref{tab:parameters} ). 

%FIXME: This explanation might not be the best.
Because of the time cost of the experiments we could only afford up to five values per parameter. We tried to keep the values used in \cite{PHTI, GideaCryptos} plus some other parameters similar to detect the sensitivity of the parameters, and one or two other options further away (see Table \ref{tab:parameters}).

%We tried to keep some values similar to the ones used in  \cite{PHTI, GideaCryptos} plus some other parameters similar to detect the sensitivity of the values, plus one or two other options further away (see Table \ref{tab:parameters}).

With a window of $w$ trading days, for each $t$ we get a point cloud in $\R^d$
\[
Z ^ t: = \{z_t, z_ {t + 1}, \ldots, z_ {t + (w-1)} \},
\]
and obtain the 0-dimensional and 1-dimensional  persistence diagrams and their persistence landscapes. Let $\{\lambda_t\}_t$ be the time series of persistence landscapes obtained. Following \cite{PHTI, GideaCryptos} ideas, we track the $L^2$-distances between the persistence landscapes $\lambda_t$ and $\lambda_{t-T}$:
\[
M_t := ||\lambda_{t-T} - \lambda_t||_2, \quad \textrm{for each } t\geq T.
\]

%DONE: Quitar que usamos T=5 ¿Estamos haciendo esto para T=30 y 60?¿Qué pasó con T=5?
 In \cite{PHTI, GideaCryptos}, the authors used $T=1$, but here we experimented with other values of $T$; we mainly used $T=1, 5, 15, 30, 60$ to compare persistence landscapes between a trading month or two of difference. We used $w=30, 60$ just to check if moving from a month of data to two months can make the index more sensitive.
 
 %and because if we move our window just a day or a week the point clouds won't change too much. We decide to try something more extreme and compare our point cloud with the one a month or two months ago, to make sure that most of the data have changed.

%DONE: Explicar que usaremos varios valores para d, tau, w y T. Y homología de Dimensión 0 y 1.
In summary, we obtain an index $I_{(d, \tau, w, T, dim)}(\{x_t\}_t)=\{M_t \;:\; N >t> T\}$ that depends on five parameters: $d, \tau, w, T$ and, the homology dimension $dim$ (0 or 1); the exact values considered are shown on Table \ref{tab:parameters}.

\begin{table}[]
    \centering
    \begin{tabular}{c|c|c|c|c|c}
        \hline 
        \multicolumn{6}{c}{\textbf{List of parameter variations}} \\
        \hline
         parameter & d & $\tau$ & $w$ & $T$ & dimension\\
         \hline
         values & 3, 4, 5, 10 & 1, 2, 5 &  30, 60 & 1, 5, 15, 30, 60 & 0,1\\
         \hline
    \end{tabular}
    \caption{Values given to the different parameters required to compute the turbulence index}
    \label{tab:parameters}
\end{table}

%DONE: Explicar que usaremos los siguientes data-sets
We tested our turbulence index on daily log-returns of the following datasets:
\begin{itemize}
    \item S\&P500: It is a stock market index that tracks the stock performance of 500 large companies listed on stock exchanges in the United States.
    \item Russel 2000: It is a small-cap stock market index made up of the smallest 2,000 stocks in the Russell 3000 Index (benchmark of the entire U.S. stock market).
    \item S\&P/BMV IPC: Weighted measurement index of 35 stocks traded on the \textit{Bolsa Mexicana de Valores} (Mexican stock exchange).
    \item Nikkei 225: Stock market index for the Tokyo stock exchange (TSE).
\end{itemize}

%DONE: Explain the reason why we choose these datasets
We decided to analyze S\&P500 and Russel 2000 to be able to compare our results with those in \cite{GideaLandscapes}; with the difference that they studied the technology crash of 2000. Also, as we want our turbulence index to be sensitive in a more robust way, we decided to include other stock datasets to remove different biases that working with only one dataset might bring. We included Nikkei and S\&P/BMV IPC to help us remove some bias against location and economy size.

%FIXME: We want to determine the best parameters for our index, testing it with different datasets.
%FIMXE: we would want that our turbulence index detects early warning signals in finantial crash in general.
As we want our turbulence index to be able to detect early warning signals of financial crashes, we focused first on adjusting the index to the 2020 stock market crash on February 20, 2020, due to the COVID-19 pandemic, when a price drop of 33\% started until March 23. We computed a total of 240 variations of the turbulence index $$\{I_{(d, \tau, w, T, dim)}(dataset)\}_{(d, \tau, w, T, dim)}$$ near COVID-19 market crash (14 months before and 3 months after February 20, 2020), where $(d, \tau, w, T, dim)$ varies over each option in Table \ref{tab:parameters} and  $dataset$ can be any log-returns time series of the four datasets mentioned above (S\&P500, Russel 2000, S\&P/BMV IPC, Nikkei 225); a total of $240 \times 4 = 960$ time-based indices.

\begin{figure}
    \centering
    \includegraphics[width=\linewidth]{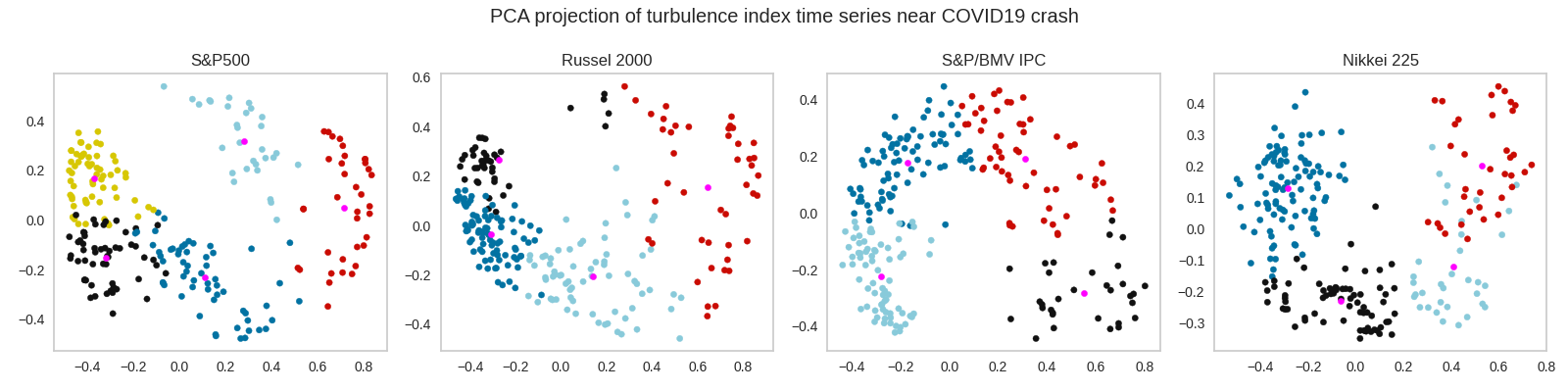}
    \caption{The 2-dimensional PCA projection of the 240 turbulence indices near COVID crash (February 20, 2020). Indices within a cluster are colored the same; pink dots represent the cluster's centroids.}
    \label{fig:pca}
\end{figure}

These indices differ a lot from one another, as it is shown in Figure \ref{fig:pca}. These plots were built cropping all indices $I_{(d, \tau, w, T, dim)}(dataset)$ to the same length and normalizing them to compare their shape and finally taking the two-dimensional PCA projections of the resulting series. As we can observe, some indices are pretty close, but many are far away. Different parameter configurations create different pattern shapes. 

\begin{figure}
    \centering
    \includegraphics[width=\linewidth]{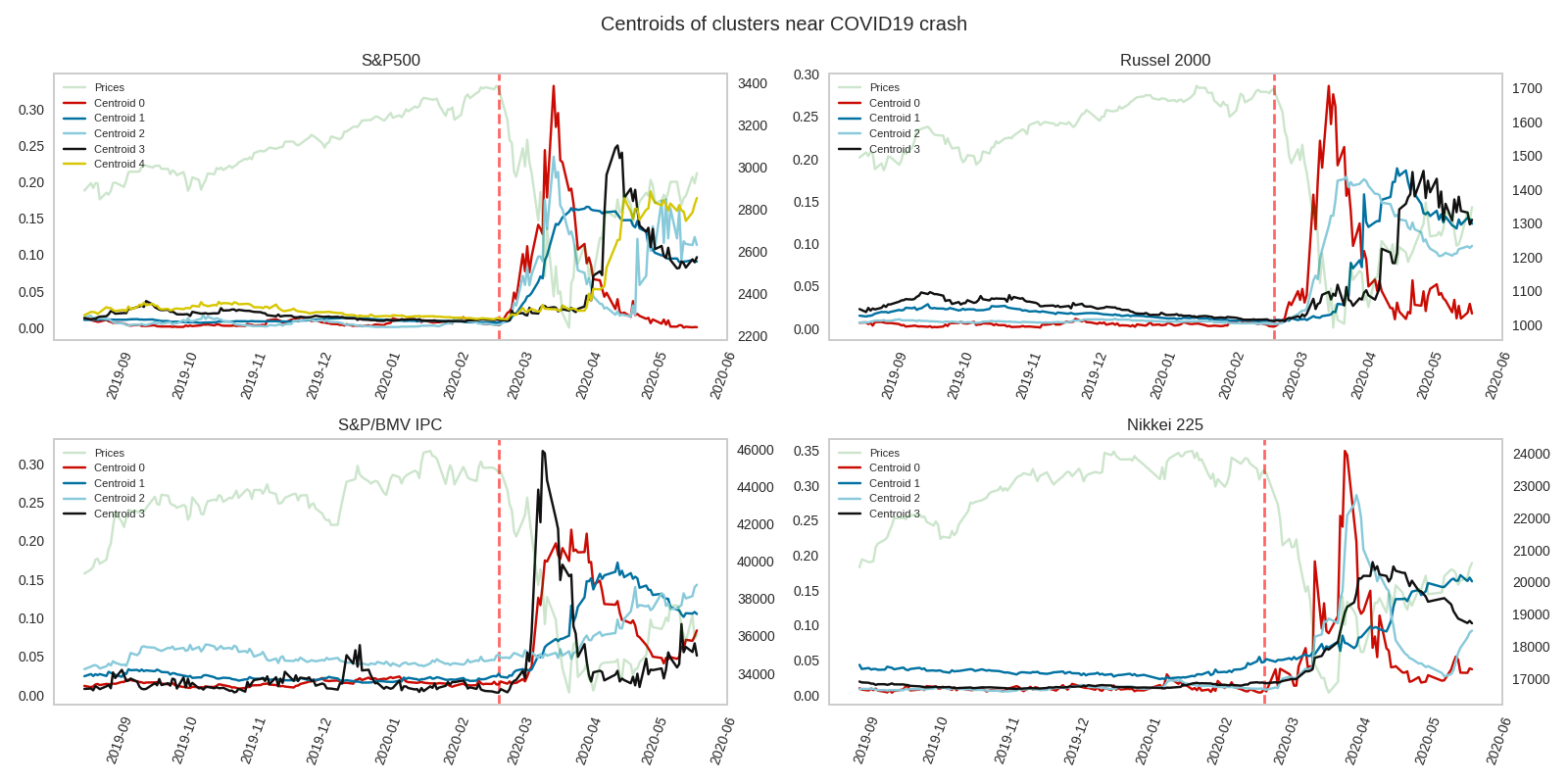}
    \caption{The graph of each turbulence-indices cluster's centroid computed through K-means. Each color corresponds to a cluster of the same color in Figure \ref{fig:pca}. The right y-axis corresponds to the price (graph in green) of the corresponding financial time series, while the left y-axis corresponds to the values of the turbulence index. The red-dotted-vertical lines represent the beginning of the COVID crash, i.e., February 20, 2020.}
    \label{fig:centroids}
\end{figure}

%DONE: Explicar cómo se hicieran los clusters.
It is natural to ask what parameter configuration better captures the risk of market crashes. To get a comprehensive description of the possible \emph{shapes} of \\$I_{(d, \tau, w, T, dim)}(dataset)$, we computed the K-means algorithm over all the normalized indices. We used the elbow rule for each dataset to decide the number of clusters to run K-means into. The final clusters are shown in Figure \ref{fig:pca}; the centroids of the clusters are shown in Figure \ref{fig:centroids}, which serve as a nice representative index of each cluster.
%The final clusters are shown in Figure \ref{fig:pca}; we obtained a representative for each cluster by taking the K-means centroids of each cluster. We plotted the centroids in Figure \ref{fig:centroids}.

%DONE: Explain the decision to choose the centroids.
%We believe that the most desired shape for a turbulence index that detects the  COVID crash is one with a prominent peak close to the beginning of the prices' fall. Also, it must return close to normal after the prices start increasing back again. 

We believe that the most desired shape for a turbulence index that detects a financial crash is one with a prominent peak close to the beginning of the prices' fall and that remains stable in non-turbulence times. For each of the centroids depicted in Figure \ref{fig:centroids} we chose the one that has that desired shape. This was the centroid 0 for S\&P500, Russel 2000, and Nikkei 225; and the centroid 3 for S\&P/BMV IPC; we call these selections the `good' centroids and the corresponding clusters the `good' clusters. There were only 14 parameter configurations $c = (d, \tau, w, T, dim)$ such that $I_c(dataset)$ was in the good cluster for all datasets. For each of them, we computed its distance from the good centroid; we show the result in Table \ref{tab:best_configurations}.

%DONE: Explicar cómo se eligieron los mejores.

\begin{table}[]
    \centering
    \begin{tabular}{l|c|c|c|c}
\hline
\multicolumn{5}{c}{\textbf{Best configurations}} \\
\hline
$(d, \tau, w, T, dim)$ &	S\&P500	& Russel 2000	& S\&P/BMV IPC & Nikkei 225\\
\hline
$(4 ,2 ,60 ,5 ,0)$ & 0.358248 & \textbf{0.435222} &0.422185 &0.395897  \\
$(5 , 2 , 60 , 5 , 0)$ & 0.325145  & 0.432211  & \textbf{0.458128}  & 0.391419 \\
$(10 , 1 , 60 , 5 , 0)$ & 0.243123  & 0.394079  & 0.410875  &\textbf{0.473084} \\
$(3 , 1 , 60 , 5 , 0)$ & 0.372211  & \textbf{0.486730}  & 0.393950  & 0.471953  \\
$(3, 5 , 60 , 5 , 0)$ & 0.264082  & 0.425575  & \textbf{0.505006}  & 0.365352  \\
$(10 , 1 , 60 , 1 , 0)$& 0.422799  & 0.415909  & 0.438392  & \textbf{0.509998}  \\
$(10, 2 , 60 , 1 , 0)$ & \textbf{0.596849}  & 0.509674  & 0.551878  & 0.500585 \\
$(5 , 1 , 60 , 1 , 0)$ & 0.620979  & \textbf{0.627667}  & 0.531569  & 0.607321 \\
$(4 , 1 , 60 , 1 , 0)$ & 0.621053  & 0.637028  & 0.593142  & \textbf{0.642688} \\
$(5 , 2 , 60 , 1 , 0)$ & \textbf{0.682100}  & 0.630856  & 0.569481  & 0.600423 \\
$(3 , 1 , 60 , 1 , 0)$ & \textbf{0.696283}  & 0.622632  & 0.617384  & 0.652061 \\
$(4 , 2 , 60 , 1 , 0)$ & \textbf{0.707310}  & 0.660973  & 0.514964  & 0.579040 \\
$(3 , 2 , 60 , 1 , 0$ & \textbf{0.715324}  & 0.617745  & 0.644957  & 0.542285 \\
$(3, 5 , 60 , 1 , 0)$ & 0.577607  & 0.570889  & \textbf{0.754845}  & 0.531487 
\end{tabular}
    \caption{List of parameters $c = (d, \tau, w, T, dim)$ such that $I_c(dataset)$ appeared as part of a good cluster for every dataset. On every column, we show the distance from  $I_c(dataset)$ to the centroid of the corresponding cluster. We mark in boldface the maximum value among the four distances.}
    \label{tab:best_configurations}
\end{table}

%DONE: Explicar que se hace un experimiento para verificar qué tan bien funcionan los mejores en otros crashes. Probablemente poner que los parámetros se ajustan mejor a los tiempos de lo mercados 

To test if the configurations from Table \ref{tab:best_configurations} are good at alerting from possible crashes in general, we tried them in a completely different set of past crashes. We worked with the following events:
\begin{itemize}
    \item Black Monday crash (see \cite{blackMondayReference} ) that occurred on October 19, 1987. 
    \item The dot-com bubble from the late 90s when many online companies failed and shut down (see \cite{dotComBubble})
    \item The United States bear market of 2007–08 is a 17-month period when the stock market lost more than 50\% of its value. 
\end{itemize}

All the above crashes are very different in nature than the COVID crash. Black Monday and dot-com are economic crashes classified as bubbles, while the 2007-08 crash is a long fall period. 

%By the way, we don't expect our index to behave the same way as in COVID crisis. But we do expect some level of sensitivity.

We computed the fourteen indices given by the parameters from Table \ref{tab:best_configurations} on each previous crash. We used the same datasets as before but moved our attention to a specific crash date. Not all datasets go back to 1987; Table \ref{tab:datasets_vs_crache} shows what crash dates appear on each dataset.

\begin{table}[]
    \centering
    \begin{tabular}{l|c |c | c }
    \hline
    \multicolumn{4}{c}{\textbf{Crashes that appear on each dataset}} \\
    \hline
     Dataset/Crash & Black Monday & Dot-com Bubble & 07–09 Bear Market   \\
    \hline
     S\&P500 &  \checkmark & \checkmark & \checkmark\\
     Russel 2000 & & & \checkmark \\
     S\&P/BMV IPC & \checkmark & \checkmark & \checkmark \\
     Nikkei 225 & & & \checkmark
    \end{tabular}
    \caption{Crash dates analyzed per dataset}
    \label{tab:datasets_vs_crache}
\end{table}

\begin{figure}
    \centering
    \includegraphics[width=\linewidth]{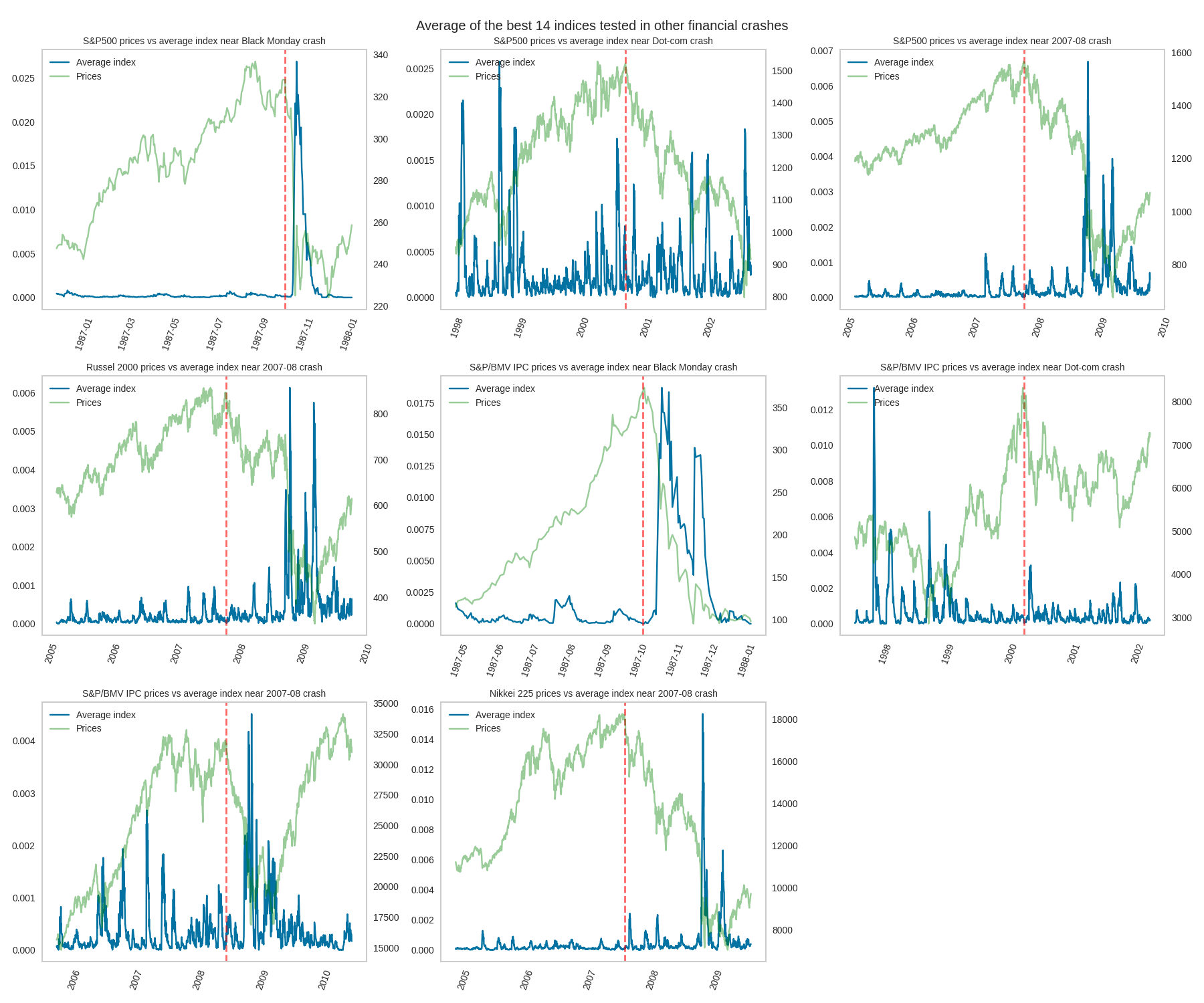}
    \caption{The average index of the fourteen normalized indices given by the parameters from Table \ref{tab:best_configurations} per crash event and dataset from Table \ref{tab:datasets_vs_crache}. The right y-axis corresponds to the price graph in green of the corresponding financial time series, while the left y-axis corresponds to the values of the average index in blue. The red-dotted-vertical lines represent the beginning of the corresponding crash.}
    \label{fig:other-crashes}
\end{figure}

%To present a general idea of the behavior of these fourteen indices, we compute the average index among them and plot the resulting average index per crash event and dataset from Table \ref{tab:datasets_vs_crache}. The final plots are presented in Figure \ref{fig:other-crashes}.
To present a general idea of the behavior of these fourteen indices, we computed the average index among the fourteen normalized indices. In Figure \ref{fig:other-crashes}, we see the plots of the resulting average index per crash event and dataset from Table \ref{tab:datasets_vs_crache}.

We can observe that for all events except two, the average index shows a prominent peak after the crash started, being the Black Monday on the S\&P500 dataset the one on which the peak fires closer to the actual crash. Observe that, in general, the position of the peaks is parallel to a sudden drop in prices, which leads us to believe that the proposed index is sensitive to this behavior. 

Cases like the dot-com crash in S\&P500 dataset have a peak that is not as prominent as the ones that occurred before and after, which can be explained by the fact that dot-com crash is a slow drop that occurred during two years \cite{dotComBubble}. In our experiments, the index only looked at the prices up to two months; so, if we want to improve the sensitivity to slow drops, we must increase the sliding window size $W$. A similar phenomenon can be observed in the 2007-08 crash on all the datasets. The index fires a peak only after a year when a sudden drop happened. 

All code and data we used can be found in a GitHub repository \cite{Miguel2022}.

\section{Conclusions} \label{Conclusions}

%FiXME: Ajustarlo a los resultados de la sección anterior.
We reviewed some efforts that have been made to understand financial markets via TDA \cite{GideaNetworks, GideaLandscapes, GideaCryptos, PHTI}, and proposed a new turbulence index based on persistence landscapes of 0 and 1-dimensional persistence diagrams. In our experiments, we found that this index shows prominent peaks near sudden price drops in financial time series caused by market crashes. The calculations of the persistent homology were done using the TDA Python library Giotto-TDA \cite{tauzin2020giottotda}. Applications of TDA on financial markets were in their early stages of development at the time this research was done, as \cite{GideaCryptos} mentioned. Those applications have been focused on detecting early warning signals of critical transition in financial markets (crashes of markets). The results are promising and indicate that there is room for collaboration between academic research and the financial industry to explore TDA insights into financial markets. 

\section*{Acknowledgements}

This research was undertaken as part of a research fellowship given by CONAHCyT through the \textit{`Sistema Nacional de  Investigadores'} sponsored by Professor José-Carlos Gómez-Larrañaga (CIMAT-Mérida). Professor Jesús Rodríguez Viorato is part of the program \emph{Investigadores por M\'exico} from CONAHCyT. 

% I am thankful to Professor José-Carlos Gómez-Larrañaga (CIMAT) for giving me the opportunity to accomplish this research and learn more about the behavior of financial markets through new tools and points of view, and also thankful to Professor Jesús Rodríguez Viorato for his patience in reviewing this paper and contributing with his ideas and hard work.

\bibliographystyle{plain}  
\bibliography{main}

\end{document}